\documentclass[twocolumn,nofootinbib,preprintnumbers,amsmath,amssymb,superscriptaddress,aps]{revtex4}

\usepackage{graphicx,subfigure,multirow}
\usepackage{longtable}
\usepackage{dcolumn}% Align table columns on decimal point
\usepackage{bm}% bold math
\usepackage[all]{xy}
\usepackage{color}
\usepackage{amsmath}
\usepackage[usenames,dvipsnames]{xcolor}
\usepackage[unicode=true,pdfusetitle,
 bookmarks=true,bookmarksnumbered=false,bookmarksopen=false,citecolor=Turquoise,
 breaklinks=false,pdfborder={0 0 1},backref=false,colorlinks=true,pdfpagemode=FullScreen]
 {hyperref}

\usepackage{ulem}
\makeatletter

\newcommand{\fmslash}[2][0mu]{%
  \mathchoice
    {\fmsl@sh\displaystyle{#1}{#2}}%
    {\fmsl@sh\textstyle{#1}{#2}}%
    {\fmsl@sh\scriptstyle{#1}{#2}}%
    {\fmsl@sh\scriptscriptstyle{#1}{#2}}}
\newcommand{\fmsl@sh}[3]{%
  \m@th\ooalign{$\hfil#1\mkern#2/\hfil$\crcr$#1#3$}}
\makeatother

\newcommand{\lsim}{{\;\raise0.3ex\hbox{$<$\kern-0.75em\raise-1.1ex\hbox{$\sim$}}\;}}
\newcommand{\gsim}{{\;\raise0.3ex\hbox{$>$\kern-0.75em\raise-1.1ex\hbox{$\sim$}}\;}}

\newcommand{\beq}{\begin{equation}}
\newcommand{\eeq}{\end{equation}}
\newcommand{\bea}{\begin{eqnarray}}
\newcommand{\eea}{\end{eqnarray}}
\mathchardef\minus="002D

\usepackage{color}

\def\beq{\begin{equation}}
\def\eeq{\end{equation}}
\def\bea{\begin{eqnarray}}
\def\eea{\end{eqnarray}}

\begin{document}
\title{Search for Boosted Dark Matter at ProtoDUNE}

\author{Animesh Chatterjee}
\email{animesh.chetterjee@uta.edu}
\affiliation{Department of Physics, The University of Texas at Arlington, Arlington, TX, USA}
\author{Albert De Roeck}
\email{Albert.de.Roeck@cern.ch}
\affiliation{CERN, Geneva, Switzerland}
\author{Doojin Kim}
\email{doojin.kim@cern.ch}
\affiliation{Theoretical Physics Department, CERN, Geneva, Switzerland}
\author{Zahra Gh.~Moghaddam}
\email{z.gh.moghaddam@gmail.com}
\affiliation{CERN, Geneva, Switzerland}
\author{Jong-Chul Park}
\email{jcpark@cnu.ac.kr}
\affiliation{Department of Physics, Chungnam National University, Daejeon 34134, Republic of Korea}
\author{Seodong Shin}
\email{seodongshin@yonsei.ac.kr}
\affiliation{Enrico Fermi Institute, University of Chicago, Chicago, IL 60637, USA}
\affiliation{Department of Physics \& IPAP, Yonsei University, Seoul 03722, Republic of Korea}
\author{Leigh H.~Whitehead}
\email{leigh.howard.whitehead@cern.ch}
\affiliation{CERN, Geneva, Switzerland}
\author{Jaehoon Yu}
\email{jaehoon@uta.edu}
\affiliation{Department of Physics, The University of Texas at Arlington, Arlington, TX, USA}

\preprint{
\begin{minipage}[b]{1\linewidth}
\begin{flushright}
CERN-TH-2018-047 \\
EFI-18-3\\
\end{flushright}
\end{minipage}
}

\begin{abstract}
We propose the first experimental test of the inelastic boosted dark matter hypothesis, capitalizing on the new physics potential with the imminent data taking of the ProtoDUNE detectors.
More specifically, we explore various experimental signatures at the cosmic frontier, arising in boosted dark matter scenarios, i.e., relativistic, inelastic scattering of boosted dark matter often created by the annihilation of its heavier component which usually comprises of the dominant relic abundance.
Although features are unique enough to isolate signal events from potential backgrounds, vetoing a vast amount of cosmic background is rather challenging as the detectors are located on the ground.
We argue, with a careful estimate, that such backgrounds nevertheless can be well under control by performing dedicated analyses after data acquisition.
We then discuss some phenomenological studies which can be achieved with ProtoDUNE, employing a dark photon scenario as our benchmark dark-sector model.
\end{abstract}

\maketitle

\section{Introduction}

The Deep Underground Neutrino Experiment (DUNE)~\cite{Acciarri:2016ooe} is projected to be in operation in 2024 and it will cover a broad physics program including precision measurements of neutrino oscillations, CP phase measurement in the lepton sector, and possibly explorations of new physics at both the intensity and cosmic frontiers, thanks to high intensity proton beams and the large mass detectors located about 1.5\,km underground at the Sanford Underground Research Facility in South Dakota, USA. 
There will be a total of four 10\,kt fiducial mass far-detector modules based on liquid Argon time projection chamber (LArTPC) technology with two initially and extending to four within a few years and a near-detector~\cite{Acciarri:2016ooe}.

For the successful operation of the DUNE experiment with kiloton-scale LArTPC detectors, a prototype of DUNE called ProtoDUNE~\cite{Abi:2017aow,Agostino:2014qoa} was planned and is under construction at CERN, anticipating the initial operation from September 2018.
The two ProtoDUNE detectors use different technologies, single-phase (SP)~\cite{Abi:2017aow} and dual-phase (DP)~\cite{Agostino:2014qoa} LArTPCs, both of which may be adopted as the DUNE far-detector, and will test the long-term stability and operation of these detectors, act as an engineering proof-of-principle for scalability, and calibrate beam and cosmic-ray responses.

While these tasks take the highest priority for the detectors, we ask whether there are physics opportunities at ProtoDUNE, in particular, considering a large active volume of 720 tons (420 tons for SP and 300 tons for DP)~\cite{Abi:2017aow,Agostino:2014qoa} and high-performance LArTPC detectors proven at Argon Neutrino Test (ArgoNeuT)~\cite{Anderson:2012vc}, Imaging Cosmic And Rare Underground Signals (ICARUS)~\cite{Antonello:2015lea}, and Micro Booster Neutrino Experiment (MicroBooNE)~\cite{Acciarri:2016smi}.
However, cosmic backgrounds will be formidable because the ProtoDUNE detectors are installed on the Earth's surface, and as a result, any signals of interest could be buried in such backgrounds.

In this paper, after performing a careful estimation, we argue that the cosmic backgrounds can be well controlled by dedicated event selections at the analysis stage and possibly, but less crucially, the addition of an efficient cosmic ray tagging apparatus.
This opens up the unexpected potential for cosmic frontier physics opportunities at ProtoDUNE and thus for DUNE.
Such a potential is indeed further advocated, as the ProtoDUNE detectors are now planned to take data for cosmic-origin signal searches.
We remark that the DUNE far-detector will start taking data $1 - 2$ years earlier than the neutrino beam becomes available and collect signals of cosmic and solar origin, thus our physics studies at ProtoDUNE will provide valuable physics input and potentially a realistic guideline for new physics searches at the DUNE far-detector.

%%%%%%%%%%%%%%%%%%%%%%%%%%%%%%%%%%%
%%%%%%%%%%%%%%%%%%%%%%%%%%%%%%%%%%%
%%%%%%%%%%%%%%%%%%%%%%%%%%%%%%%%%%%

\section{Benchmark physics scenario}

An exciting physics opportunity with ProtoDUNE is the search for dark matter (DM).
Unfortunately, the conventional DM search via its non-relativistic scattering is not accessible because the expected threshold energy for electron/nucleon recoil ($\sim 30$ MeV)~\cite{Acciarri:2015uup} is far beyond the typical energy deposit resulting from the ordinary DM scattering.
By contrast, typical energy deposits in association with a relativistic scattering of boosted DM readily surpass such a threshold, which renders ProtoDUNE as an ideal detector in the search for boosted DM with its relativistic scattering signatures.

\begin{figure*}[t]
\centering
\includegraphics[width=14cm]{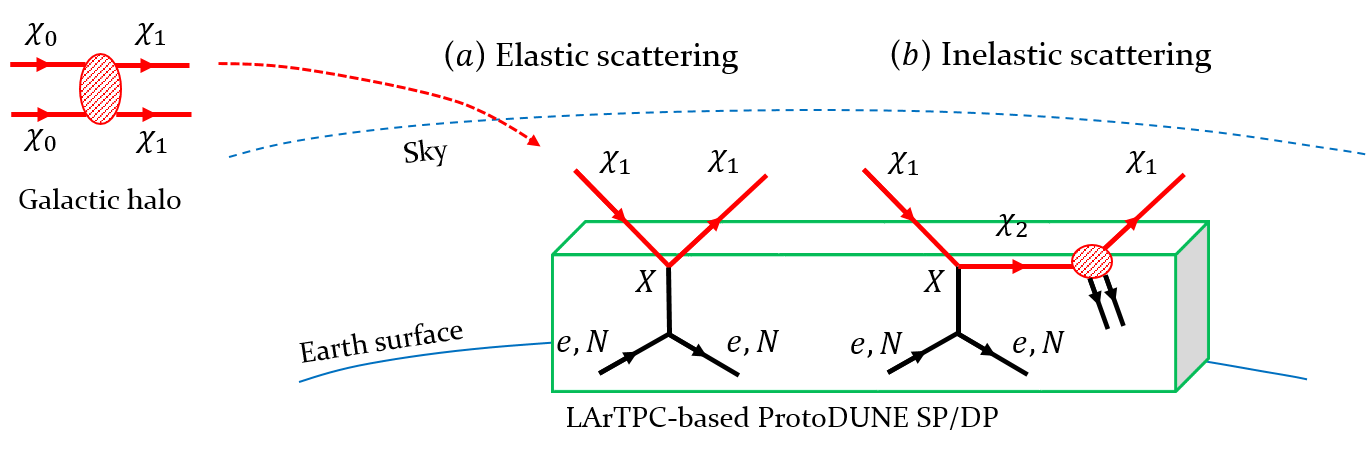}
\caption{\label{fig:scenario} The processes under consideration with the ProtoDUNE detectors. Secondary particles in $(b)$ are possibly visible on top of the visible target recoil.}
\end{figure*}

A possible mechanism to create relativistic DM in the current universe is the boosted dark matter (BDM) scenario~\cite{Agashe:2014yua} which hypothesizes two stable DM species: the heavier $\chi_0$ and the lighter $\chi_1$.
Their overall relic abundance is determined by the ``assisted freeze-out'' mechanism~\cite{Belanger:2011ww}, and in typical cases the heavier (lighter) becomes the dominant (negligible) relic as it has {\it in}direct coupling to the Standard Model particles through the lighter component.
Therefore, in the current universe, a pair of $\chi_0$ annihilate into a pair of $\chi_1$ in the galactic halo.

The mass gap between the two species allows $\chi_1$ to acquire a large boost factor and induce relativistic scattering signatures in terrestrial detectors.
FIG.~\ref{fig:scenario} shows two such possible processes.
The process on the left corresponds to the ordinary elastic scattering with a visible target recoil~\cite{Agashe:2014yua,Berger:2014sqa,Kong:2014mia,Alhazmi:2016qcs} (henceforth called {\it e}BDM).
The process on the right assumes a non-minimal dark-sector scenario allowing the transition to a heavier unstable state ($\chi_2$) which subsequently disintegrates back to $\chi_1$ together with possibly visible secondary particles in addition to the primary target recoil~\cite{Kim:2016zjx,Giudice:2017zke} (henceforth called {\it i}BDM).
We define the masses of the dark-sector particles $\chi_i$ as $m_i$ for $i = (0,1,2)$.

To investigate signal detection prospects at any given experiment, it is crucial to estimate the total flux of the incoming boosted $\chi_1$, which is given by~\cite{Agashe:2014yua}
\begin{align}
\mathcal{F} &= 1.6 \times 10^{-4}\, {\rm cm}^{-2}{\rm s}^{-1} \, \times \left( \rm GeV/m_{0}\right)^2 \nonumber \\
&\hspace{0.5cm} \times  \frac{\langle \sigma v \rangle_{0 \to 1}}{ 5\times 10^{-26}\, {\rm cm}^3{\rm s}^{-1}} \, ,
\label{eq:flux}
\end{align}
where the reference value for $\langle \sigma v \rangle_{0 \to 1}$, the velocity-averaged annihilation cross section of $\chi_0 \chi_0 \to \chi_1 \chi_1$, corresponds to an observed DM thermal relic density~\cite{Agashe:2014yua,Belanger:2011ww} assuming $\chi_0$ and $\bar{\chi}_0$ are distinguishable.
Considering the fiducial volume of the ProtoDUNE detectors and assuming 2-year data collection at 50\% duty factor (i.e., $3\times 10^7$\,s), we find that ProtoDUNE is capable of probing models with $m_0$ in the range $\mathcal{O}(30 \hbox{\,MeV})- \mathcal{O}(10 \hbox{\,GeV})$.

While numerous DM models conceiving the aforementioned signatures are available, we employ the following dark photon scenario throughout this paper for illustration:
\bea
\mathcal{L} \supset &-&\frac{\epsilon}{2} F_{\mu\nu}X^{\mu\nu}\nonumber \\
&+&g_{11}\bar{\chi}_1\gamma^{\mu}\chi_1 X_\mu +\left( g_{12}\bar{\chi}_2\gamma^{\mu}\chi_1 X_\mu + {\rm h.c.}
\right). \, \label{eq:lagrangian}
\eea
The first term describes the usual kinetic mixing between the field strength tensors $F_{\mu\nu}$ for the ordinary Standard Model photon and $X_{\mu\nu}$ for the dark photon $X$ by the amount $\epsilon$.
The second (third) operator describes the flavor-conserving (flavor-changing) neutral current responsible for elastic (inelastic) scattering.
Given this scenario, we expect three types of signal events in association with electron recoil,
that is, 
\begin{itemize} \itemsep1pt \parskip0.5pt \parsep0.5pt
\item[i)] {\it e}BDM: $\chi_1 e^- \rightarrow \chi_1 e^-$, 
\item[ii)] prompt {\it i}BDM: $\chi_1 e^- \rightarrow \chi_2 e^- \rightarrow \chi_1 X (\rightarrow e^-e^+) e^-$, 
\item[iii)] displaced {\it i}BDM: $\chi_1 e^- \rightarrow \chi_2 e^- \rightarrow \chi_1 e^-e^+ e^-$.
\end{itemize}
Here we divide the {\it i}BDM case into two subcategories, whether or not the secondary $e^+e^-$ pair comes from the decay of a long-lived particle $X$ (prompt) or $\chi_2$ (displaced).

Searches for similar signatures can be done at fixed target experiments, with active production of relativistic dark matter. 
An ample amount of literature has been focusing on the elastic scattering channel, e.g. see Refs.~\cite{Batell:2009di,deNiverville:2011it}.
By contrast, phenomenological consideration to the inelastic scattering channel is being increasingly made~\cite{Pospelov:2013nea,Izaguirre:2014dua,Izaguirre:2017bqb}. 

%%%%%%%%%%%%%%%%%%%%%%%%%%%%%%%%%%%
%%%%%%%%%%%%%%%%%%%%%%%%%%%%%%%%%%%
%%%%%%%%%%%%%%%%%%%%%%%%%%%%%%%%%%%

\section{Background consideration}

We are now in the position to discuss potential backgrounds to {\it i}BDM signals.
In general, it is hard for conventional cosmic-origin events to mimic the signal due to several distinguishing  features. 
Nevertheless, we consider plausible scenarios that could  give rise to potential background events since both SP and DP detectors are placed on the ground and exposed to a high cosmic-ray rate, followed by discussions on useful background rejection strategies.

%--!
\begin{figure*}%[htp!]
\centering
\includegraphics[width=0.8\linewidth]{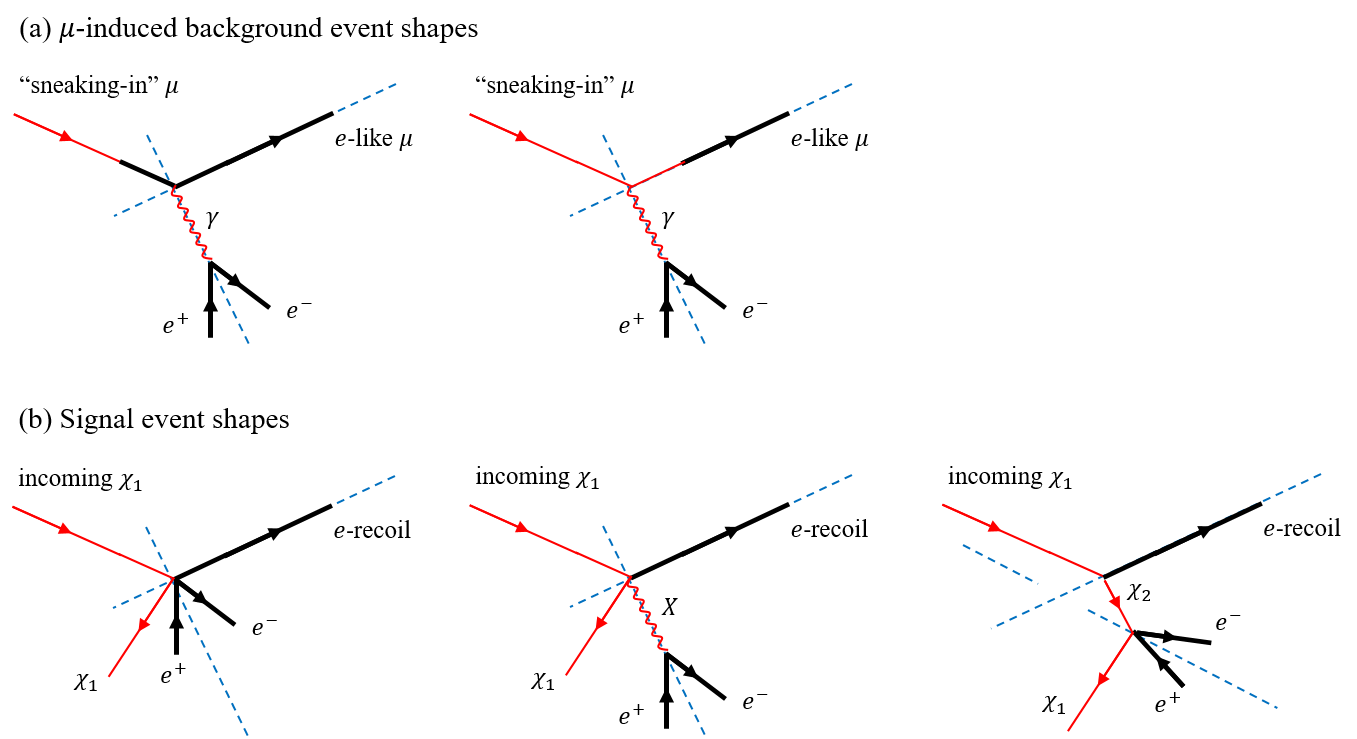}
\caption{\label{fig:eventshape} Possible event shapes of muon-induced background (upper panel) and $i$BDM signal (lower panel). The red solid lines imply that particles leave no visible tracks, whereas the black solid lines represent $e^\pm$ or $e$-like visible tracks. The blue dashed lines extend the momentum directions of the electron recoil, $e$-like muon, and $e^+e^-$ pair coming from $\gamma$ conversion or the decay of on-/off-shell dark photon $X$.}
\end{figure*}
%--!

Firstly, let us estimate the cosmic background anticipated at the ProtoDUNE detectors, separating it into low-energy cosmic rays ($30 \hbox{ MeV}\lesssim E \lesssim 400 \hbox{ MeV}$) and high-energy ones ($E \gtrsim 400 \hbox{ MeV}$).
As we will see shortly, we expect that the majority of low-energy cosmic background is suppressed, considering the (partially covered) outermost steel exoskeleton, the insulator region, and the LAr volume outside the active volume.
We further take out 35\,cm inward from the boundary of the active volumes as per DUNE conceptual design report (CDR) Vol.~IV~\cite{Acciarri:2016ooe}.
For the DP, we additionally cut away 1\,m from the top surface of LAr to offset the passive volume existing in the other sides.
These considerations reduce the fiducial volumes of the SP and DP to 300\,t and 170\,t, respectively.

Highly energetic cosmic particles such as muons, however, are not sufficiently removed even with the above-defined fiducial volumes.
Indeed, the integral intensity of vertical muons above 1\,GeV at the altitude of the actual site ($\sim400$\,m) is about 70\,m$^{-2}$s$^{-1}$sr$^{-1}$ (which negligibly differs from that at sea level~\cite{Tanabashi:2018oca}), and the muon energy spectrum below 1\,GeV is almost flat~\cite{Tanabashi:2018oca}.
We find that muons above $\sim 400$ MeV may reach the fiducial volume for both SP and DP detectors, taking into account the energy-dependent stopping power for muons in LAr (1\,m from the passive volume + 35\,cm by fiducialization)~\cite{Tanabashi:2018oca,bnl}.

Considering the flatness of the muon energy spectrum below 1\,GeV and the muon flux at 1 GeV~\cite{Tanabashi:2018oca}, we estimate $\sim 24$ m$^{-2}$s$^{-1}$sr$^{-1}$ in-between 400\,MeV and 1\,GeV.
%, hence expect about $3 - 4$\,m$^{-2}$s$^{-1}$sr$^{-1}$ in total for the detectors with their respective active volumes taken into consideration (see also Table~\ref{tab:fiducial} for their dimensions).
Hence, we conclude that $\sim60$ cosmic-ray muons per detector will enter the active volume within the 5\,ms trigger-window~\cite{Abi:2017aow} allowing for a 2.25\,ms electron drift time~\cite{Abi:2017aow}, with a range of $\pi$ steradians included.
This is within the ProtoDUNE data-recording capacity.

For the data analysis, most of the cosmic-muons will be easily recognized since they leave identifiable tracks in the detectors.
However, one may argue that a significant number of muons could still mimic signal despite small rates of missed tracks, particle misidentification, and other cases described below, purely due to the fact that the total number of muons with energies above 400 MeV is as large as $\sim 4 \times 10^{11}$ per year in each detector.
One plausible scenario to fake an {\it i}BDM signal is that a muon i) enters the fiducial volume without leaving a track, i.e., ``sneaks-in'', ii) emits a hard photon which converts into an $e^+e^-$ pair, and iii) starts to leave a visible track resulting in a signal-like event shape, iv) which appears electron-like.
While a more thorough and dedicated study on ``sneaking-in'' muons under the ProtoDUNE environment is highly desired, we can estimate the effect from a study of the muon reconstruction efficiency at the MicroBooNE detector~\cite{MicroBooNENote}. They reported that $0.09\%$ of cosmic muons are reconstructed such that tracks appear only inside the fiducial volume, with the more advanced selection scheme.
We take this as the upper limit of the ``sneaking-in'' muon probability, thus conservatively estimate the probability to be $0.1\%$.\footnote{The number $0.09\%$ resulted from 2016 data of the MicroBooNE detector. 
The corresponding number including 2017 data is even smaller, although not public yet~\cite{private}.}

The second condition can be given by a phase-space suppression factor, $\alpha/\pi\approx 2\times 10^{-3}$, with $\alpha$ being the fine structure constant.
For the rate of electron-like muon tracks, a dedicated analysis is again needed, but here we simply take a very conservative suppression factor of $10^{-2}$ based on the study reported in Ref.~\cite{Acciarri:2016sli} with the LArTPC detector of the ArgoNeuT Collaboration.
The remaining criterion iii), where the momentum direction of the $e^+e^-$ pair intersects at most with the beginning point associated with the $e$-like outgoing $\mu$ track, is hard to estimate.
But we see that if $\sim 0.6$\% of suppression power is achieved, it should be possible to have fewer than $\sim 100$ muon-induced background events per year in the two ProtoDUNE detectors.
Note, however, that this estimation is based on very conservative probabilities written in criteria i), iii), and iv).
In reality, a dedicated analysis should easily decrease  these rare possibilities by a few orders of magnitude.
Nevertheless, we show the experimental sensitivities assuming 100 background events per year in order to clarify that ProtoDUNE can probe the iBDM signals even in the  worst case scenario.
We compare the results with the sensitivities for a best case scenario (zero-background assumption)\footnote{Note also that it is possible to keep the zero-background assumption following the proposal that all the cosmic muon background events can be rejected by considering the Earth shielding effect in a surface-based detector~\cite{Kim:2018veo}.} and a two-year exposure of the detector.

It is informative to understand the many topological differences between $\mu$-induced background events and the signals.  
We display the possible event shapes of $\mu$-induced background (upper panel) and $i$BDM signal of interest (lower panel) in FIG.~\ref{fig:eventshape}.
For background events, a hard photon emission (red wavy lines), which can show a visible gap with the radiation length being $\mathcal{O}(10\,\hbox{cm})$~\cite{Wolfenstein:1977ue}, may arise either after (upper left diagram) or before (upper right diagram) the ``sneaking-in'' muon (red solid lines) begins to leave an $e$-like track.
Note that for both cases, the incoming (sneaking-in) muon, the outgoing ($e$-like) muon, and the hard photon lie on a common plane. 
In other words, the line extending the momentum direction of the $e^+e^-$ pair from the $\gamma$ conversion should meet the line extending the momentum direction of the outgoing $e$-like track, within the detector position resolution.
For signal events, three event shapes are possible. 
Firstly, if primary scattering and secondary decay take place promptly, all three electron tracks are expected to start at a single vertex point (lower left diagram).
Secondly, if $\chi_2$ decays instantly but the dark photon $X$ is long-lived (lower middle diagram), the direction extended by the total momentum of an $e^+e^-$ pair from $X$ decay should point back the beginning point associated with the electron recoil track, within the detector position resolution.
Finally, if $\chi_2$ is long-lived, it decays to $\chi_1$ and an $e^+e^-$ pair via a three-body process. 
Unlike the previous case, the line extending the momentum direction of the $e^+e^-$ pair does {\it not} intersect with that of recoil electron (lower right diagram).

Other plausible situations were considered, such as di-muon simultaneous scattering and muon-initiated deep inelastic scattering in both active and passive volumes against the {\it i}BDM signals. 
Each muon of these background events must satisfy the aforementioned criteria i) and iv) for the major background which suppress the number of background at least $10^{-5}$ for each muon.
Also, the probabilities that a single electron faking the $e^+ e^-$ and a single photon faking an electron signal are below 10\% and 7\%, respectively~\cite{Acciarri:2016sli}.
Then, taking into account the number of two simultaneously incoming muons within the detector resolution ($2.5 \cdot 10^9/{\rm yr}$) and the muons inducing the deep inelastic scattering ($1.6 \cdot 10^5/{\rm yr}$), we predict the corresponding minor background events are less than 
%and the probability of having two simultaneously incoming muons or deep inelastically scattering muon suppresses the expected number of background further.
%We hence predict 
$\sim 0.025$ and $\sim 0.11$ %interactions
per year per detector, respectively.

We have checked other high-energy cosmic particles such as electrons/positrons and pions, but their contribution is negligible because their fluxes at sea level are smaller than that for the muon by $3-4$ orders of magnitude~\cite{Tanabashi:2018oca} and their corresponding stopping powers in material are larger than that of the muon.
However, the neutron flux is only a factor of 100 less than the muon flux and not negligible.
Neutrons with GeV-range energies couple to matter via the strong force, thus they quickly break apart in material.
On the other hand, MeV-range neutrons can sneak in the detector fiducial volume and (predominantly) scatter off nuclei.
However, considering that two simultaneous ``sneaking-in'' neutrons would be required produce two $e$-like nucleon recoil tracks, we estimate that the number of the expected events is much smaller than one.

%%%%%%%%%%%%%%%%%%%%%%%%%%%%%%%%%%%
%%%%%%%%%%%%%%%%%%%%%%%%%%%%%%%%%%%
%%%%%%%%%%%%%%%%%%%%%%%%%%%%%%%%%%%

\begin{table}[t]
\centering
\begin{tabular}{c|c c}
Detector & w $\times$ h $\times$ d [m$^3$] & Active volume [t] \\
\hline
ProtoDUNE SP~\cite{Abi:2017aow} & $2 (3.6 \times 7 \times 6 )$ & 420 \\
ProtoDUNE DP~\cite{Agostino:2014qoa} & $6\times 6 \times 6$ & 300
\end{tabular}
\caption{\label{tab:fiducial} Detector specifications relevant to phenomenology in this paper.
Fiducial volumes can be inferred by taking 35\,cm inward from the boundary of the active volumes.}
\end{table}

Finally, we discuss potential atmospheric neutrino backgrounds.
The DUNE Collaboration performed a dedicated study on its event rates including oscillations in 350\,kt$\cdot$yr with a LArTPC, fully or partially contained in the detector fiducial volume~\cite{Acciarri:2015uup}.
From the fully contained electron-like sample, we estimate $\sim 40$\,yr$^{-1}$kt$^{-1}$ which may include multi-track events which can be background to the {\it i}BDM signals.\footnote{Strictly speaking, the oscillation effect at ProtoDUNE differs from that at the DUNE because they are placed in different depths below the surface.
We have explicitly checked and found that such an effect is at most $\mathcal{O}(1\%)$ for the energy scale of our interest.
Also, the MSW effect~\cite{Wolfenstein:1977ue, Mikheev:1986gs} may require a more precise estimate, which is expected to be subleading, hence beyond the scope of this paper.}
We compared this number with the Super-Kamiokande (SK) atmospheric neutrino data~\cite{Abe:2014gda} and an official {\it e}BDM analysis conducted by the SK Collaboration~\cite{Kachulis:2017nci}, and found that the number is comparable to the number of events with single-ring, $e$-like, 0-decay electrons, and 0-tagged neutrons at the SK.

For the multi-track atmospheric neutrino interactions, the DUNE CDR does not provide detailed information, but an SK study~\cite{Abe:2014gda} reported the number of multi-ring, $e$-like events from which we estimate $\sim 5$ yr$^{-1}$kt$^{-1}$.
Given the fiducial volumes of the ProtoDUNE detectors defined in Table~\ref{tab:fiducial} and the fact that extra tracks typically originating from meson decays in neutrino events can be well-identified at ProtoDUNE, the expected number of neutrino background events to the {\it i}BDM signals is negligible.\color{black}

\section{Phenomenology}

\begin{figure*}%[t]
\centering
\includegraphics[width=7.8cm]{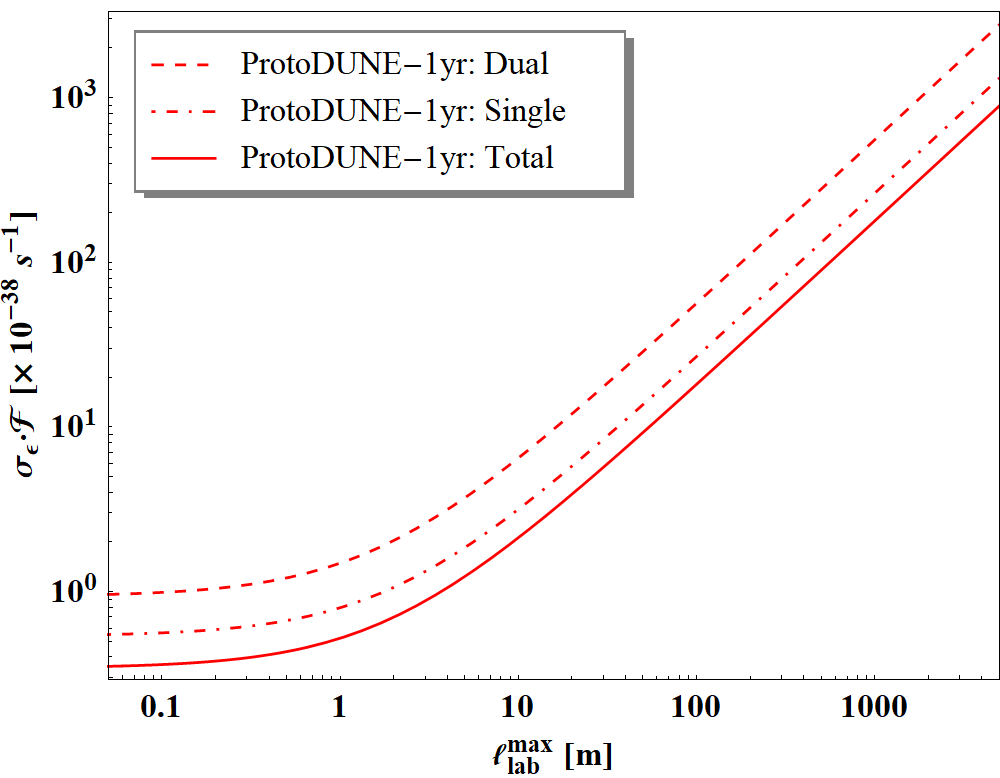} \hspace{0.5cm}
\includegraphics[width=7.8cm]{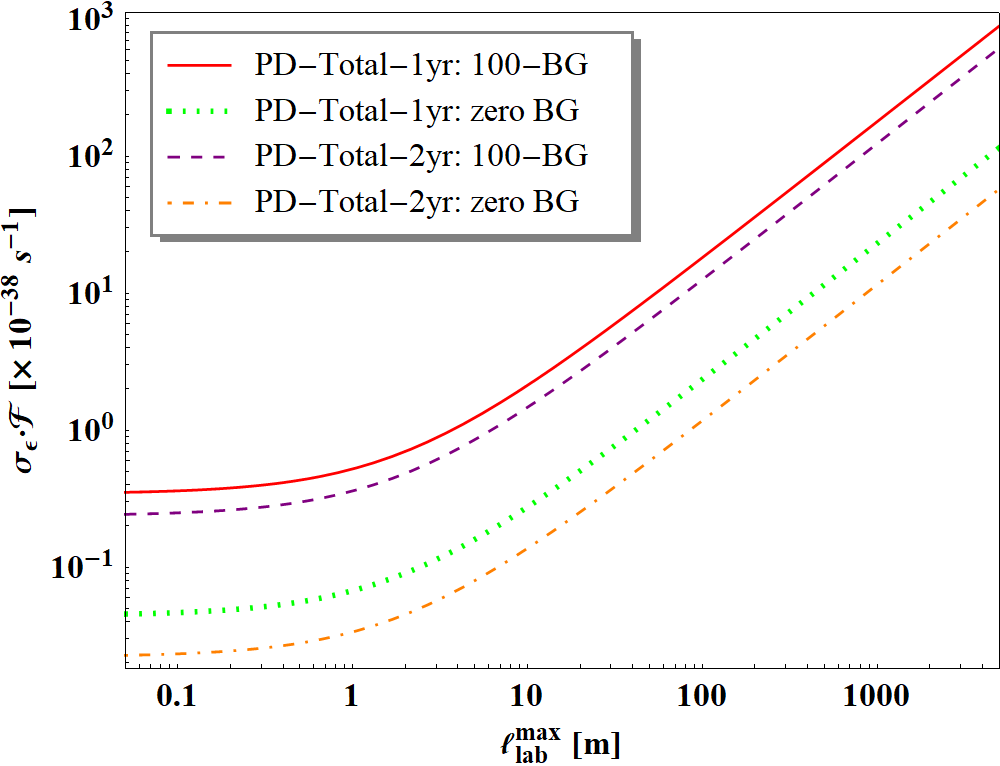} \hspace{0.5cm}
\includegraphics[width=7.8cm]{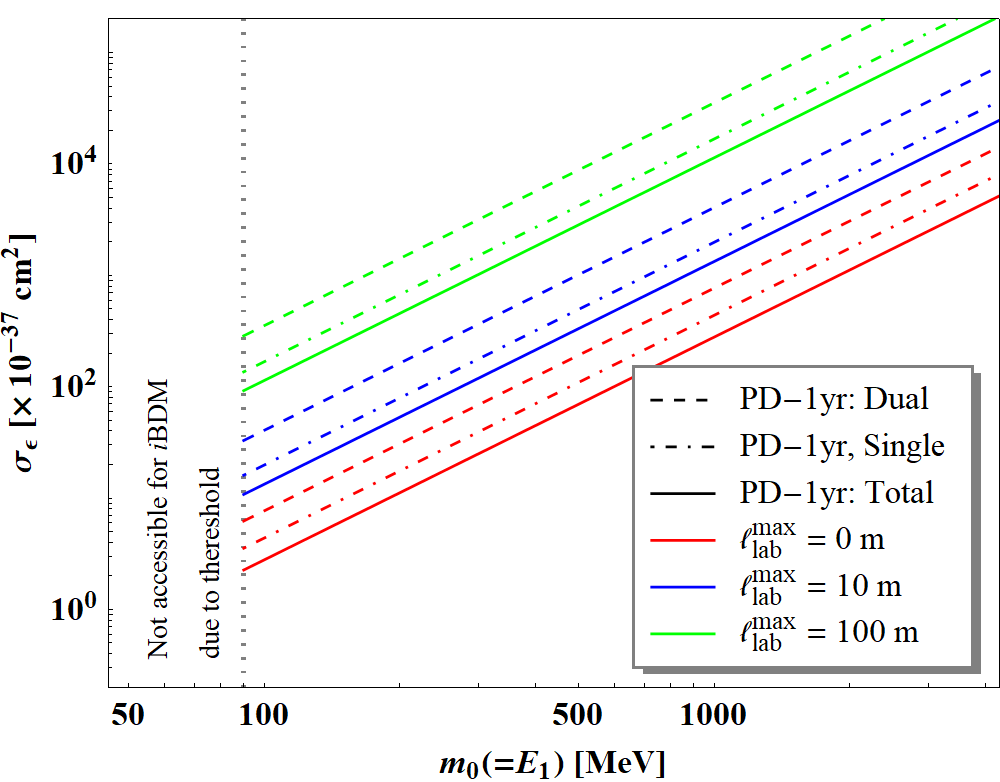} \hspace{0.5cm}
\includegraphics[width=7.8cm]{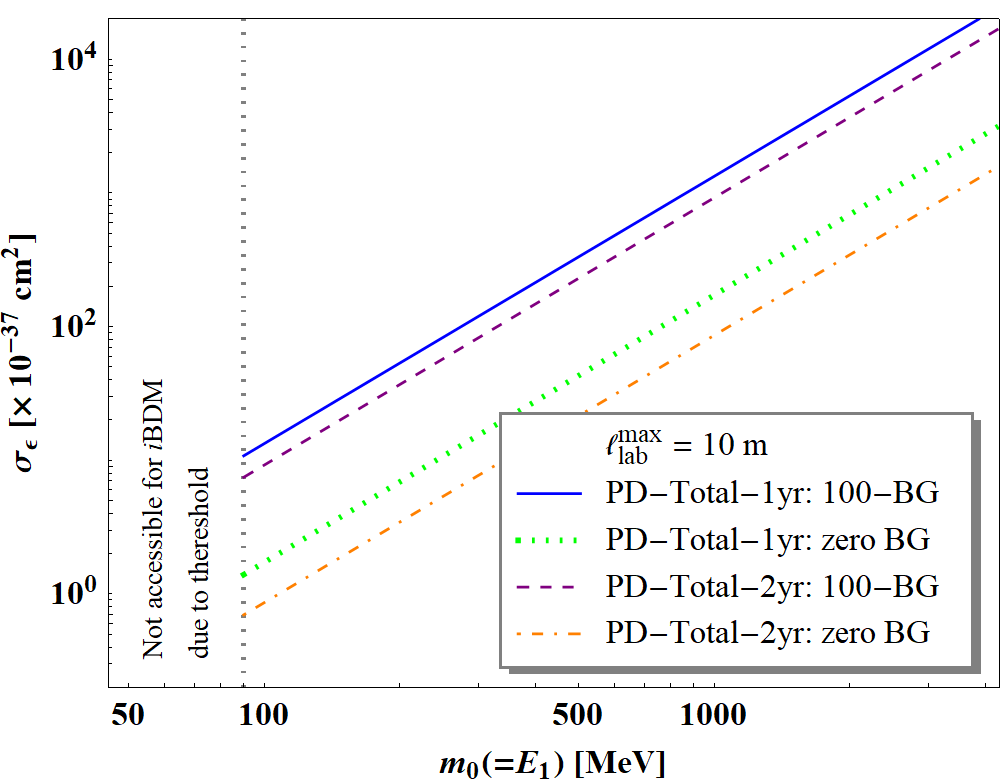}
\caption{\label{fig:sense} 
Top: Experimental sensitivities in the $\sigma_\epsilon\mathcal{F}$ vs. $\ell_{\textrm{lab}}^{\max}$ planes for the case of a displaced secondary vertex. The left panel shows the coverage in a worst case scenario assuming 100 background interactions (denoted as 100-BG) for a one-year exposure and the right panel includes the sensitivities in a best case scenario (zero-background assumption denoted as zero BG) and a two-year exposure of the detector for comparison. 
Bottom: Corresponding experimental sensitivities in $\sigma_\epsilon$ vs. $m_0(=E_1)$ for three different fixed $\ell_{\textrm{lab}}^{\max}$ values. 
}
\end{figure*}

We first discuss ways of presenting model-independent experimental reaches with respect to various physics models conceiving {\it e}BDM and/or {\it i}BDM signatures.
To this end, we consider the number of signal events, $N_{\textrm{sig}}=\sigma_\epsilon \mathcal{F} A\, t_{\textrm{exp}} N_e$, where $\sigma_\epsilon$ is the cross section of either $\chi_1 e^- \rightarrow \chi_1 e^-$ for {\it e}BDM or $\chi_1 e^- \rightarrow \chi_2 e^-$ for {\it i}BDM, $\mathcal{F}$ is the flux of $\chi_1$, $A$ is the acceptance, $t_{\textrm{exp}}$ is the exposure time, and $N_e$ is the number of target electrons inside the fiducial volume.
For the {\it i}BDM case, the branching fraction of $\chi_2 \rightarrow \chi_1 e^- e^+$ is assumed to be 1.
Note that the characteristics of the experiment determine the last two parameters ($t_{\textrm{exp}}$ and $N_e$) and refer to Table~\ref{tab:fiducial} for those of ProtoDUNE.
By contrast, the product of the first two parameters ($\sigma_\epsilon$ and $\mathcal{F}$) depends on all model parameters such as coupling constants and masses.
Finally, we assume that the acceptance $A$ (defined as 1 if the interaction is fully contained in the fiducial volume, and 0 otherwise) is determined by the distance between the primary and the secondary vertices $\ell_{\textrm{lab}}$ while all the other effects from selection criteria, threshold energy, detector response, and so on are encapsulated in the quantity $\sigma_\epsilon$.\footnote{Obviously, $A$ for $e$BDM signals is defined as 1 as all visible particles (here recoil electron only) come out of a single interaction point, i.e., no displaced vertex. }

\begin{figure*}%[t]
\centering
\includegraphics[width=7.8cm]{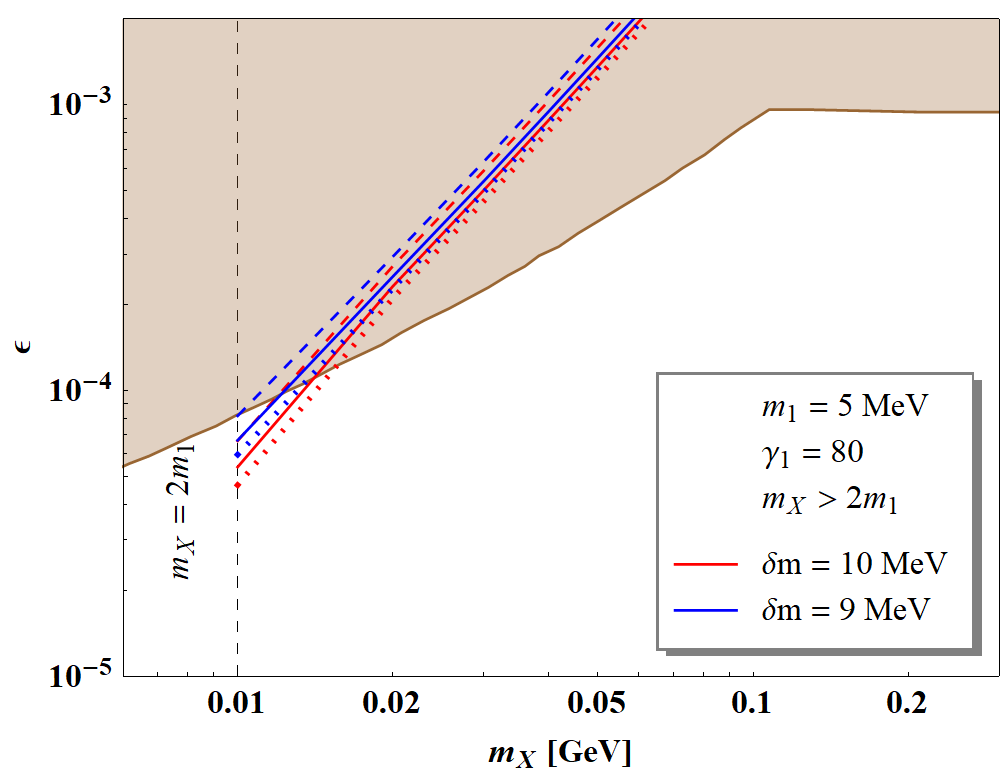} \hspace{0.5cm}
\includegraphics[width=7.8cm]{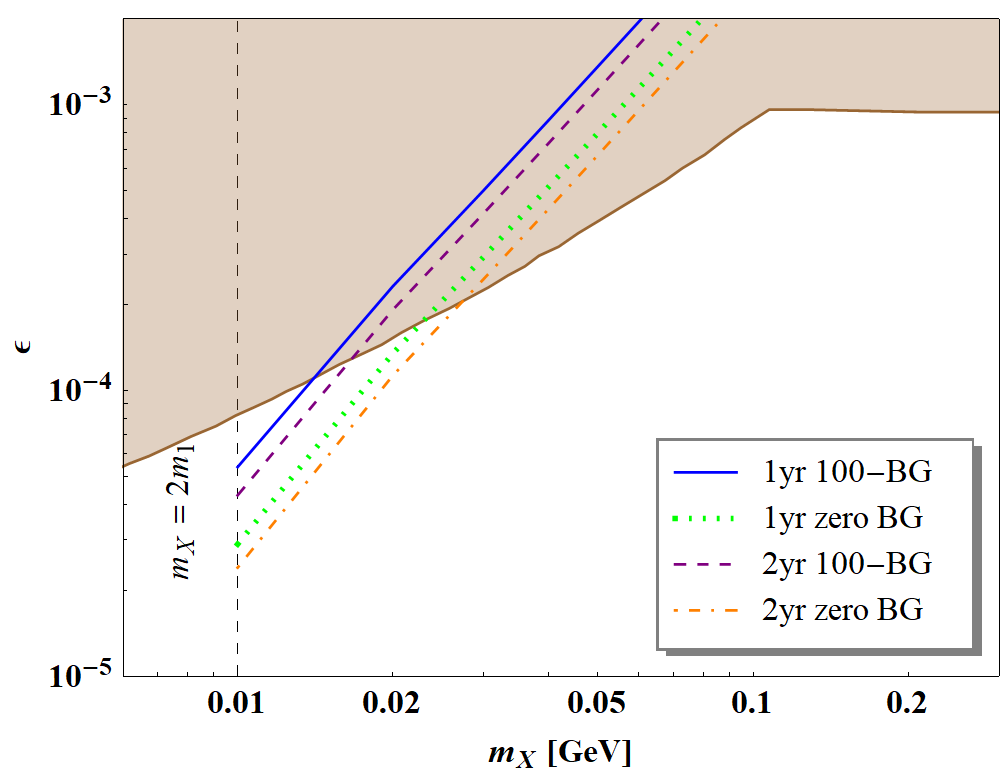} \hspace{0.5cm}
\includegraphics[width=7.8cm]{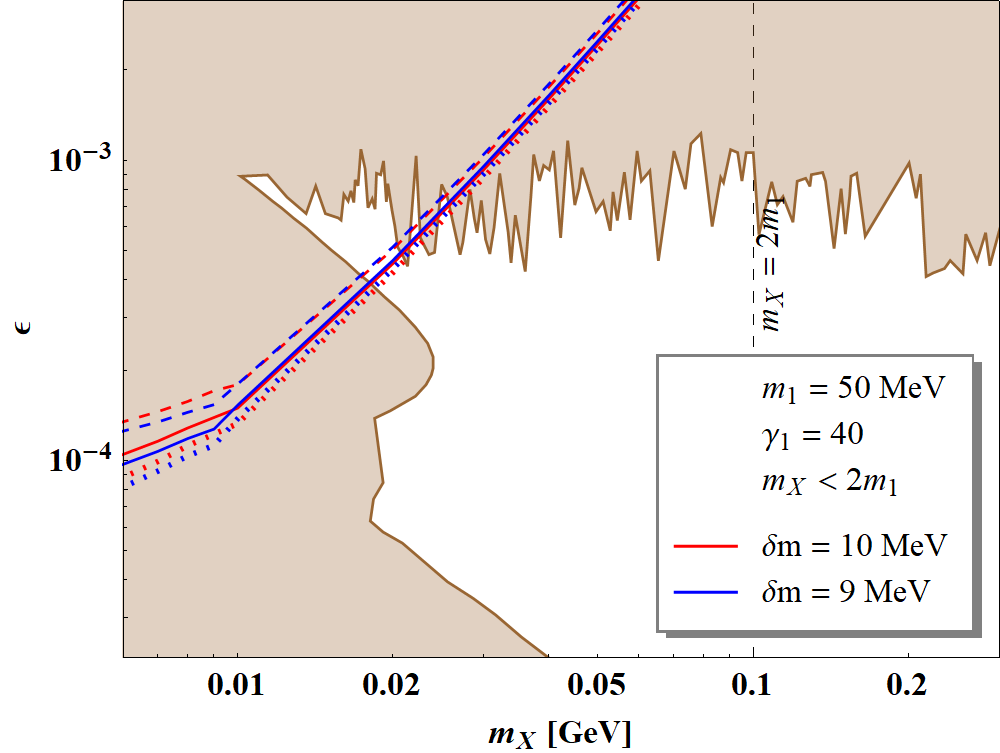} \hspace{0.5cm}
\includegraphics[width=7.8cm]{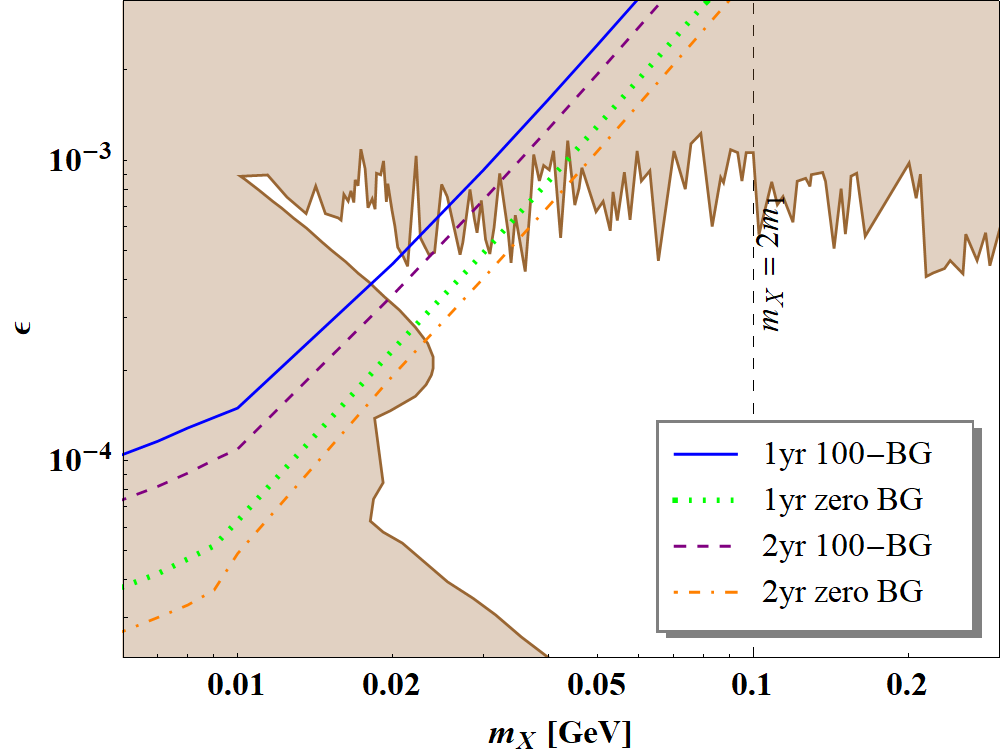}
\caption{\label{fig:darkphotonsensitivity} 
Experimental reach for one year of ProtoDUNE running in $m_X$ vs. $\epsilon$ for the cases of invisible (top panels) and visible (bottom panel) decays of the dark photon $X$. The brown shaded regions show the currently excluded parameter space, as reported in Refs. \cite{Banerjee:2017hhz} (top) and \cite{Banerjee:2018vgk} (bottom).
Just like FIG.~\ref{fig:sense}. The left panels show the coverage in a worst case scenario assuming 100 background events (denoted as 100-BG) for a one-year exposure and the right panels include the sensitivities in the best case scenario (zero-background assumption denoted as zero BG) and a two-year exposure of the detector for comparison. 
}
\end{figure*}

A possible presentation scheme is to show model-independent reaches in the plane of $\sigma_\epsilon \mathcal{F}$ vs. $\ell_{\textrm{lab}}$ which can be formally expressed as~\cite{Giudice:2017zke}
\bea
\sigma_\epsilon \mathcal{F} > \frac{N_s^{90}}{A(\ell_{\textrm{lab}})\cdot t_{\textrm{exp}}\cdot N_e}\,, \label{eq:inequality}
\eea
where the numerator $N_s^{90}$ corresponds to the 90\% C.L. upper limit of the signal events with Poisson statistics. 
$N_s^{90} = 2.3$ under a zero-background assumption which is our optimistic scenario and $N_s^{90} = 17.8$ in our worst case scenario where the number of background events is $\sim 100$.
Note that all model-dependent information is encoded in the left-hand side, whereas the right-hand side takes only experimental specifications and $\ell_{\textrm{lab}}$ which alters event-to-event. We simply follow the suggestion in Ref.~\cite{Giudice:2017zke} and exhibit the experimental sensitivity in the plane of $\sigma_\epsilon \mathcal{F}-\ell_{\textrm{lab}}^{\max}$ with $\ell_{\textrm{lab}}^{\max}$ being the maximum laboratory-frame mean decay length of a long-lived particle (here either $X$ or $\chi_2$).
We display the experimental reaches of the ProtoDUNE detectors for a one-year run period in the top panels of FIG.~\ref{fig:sense}, assuming a cumulatively isotropic $\chi_1$ flux.
The left panels show the sensitivities assuming the worst case scenario of 100 background events, while the right panels include the experimental reaches in the best case, zero-background scenario.
In the right panels, we also show the sensitivities when ProtoDUNE takes cosmic data for two years for comparison.
Note that $N_s^{90}$ takes 24.6 for the worst scenario as the total number of expected background events during 2-year data collection doubles, i.e., $\sim 200$.
Given a model having a BDM process, one can simply evaluate $\sigma_\epsilon$, $\mathcal{F}$, and $\ell_{\textrm{lab}}^{\max}$ [see Appendix B of Ref.~\cite{Giudice:2017zke} for useful formulae associated with our benchmark model in \eqref{eq:lagrangian}] to find the corresponding coordinate in the plane.
If it appears above a certain curve, the model point of interest is excluded with respect to the associated experiment/detector, and otherwise it remains an allowed point.

While this way of presentation is useful {\it per se}, a more familiar form is available.
The flux $\mathcal{F}$ is a function of the mass of the dominant relic $\chi_0$ as shown in Eq.~\eqref{eq:flux}. So, moving the flux factor in the inequality of Eq.~\eqref{eq:inequality} to the right-hand side, we are able to show the experimental sensitivities in terms of $\sigma_\epsilon$ vs. $m_0 (=E_1)$ for a given $A$ which is uniquely mapped to a value of $\ell_{\textrm{lab}}^{\max}$.
This scheme is reminiscent of the sensitivity plot in the plane of spin-independent or spin-dependent cross section vs. the mass of the dominant relic DM which is usually reported by ordinary DM direct detection experiments.
Example curves are shown in the bottom panels of FIG.~\ref{fig:sense} with three different decay lengths.
The black vertical dotted line represents the absolute lower bound for visible tri-track events due to the threshold energy of 30 MeV.
The actual lower bound may involve minor model-dependence; for a given $m_0$, it gets closer to the absolute one as $m_1$ becomes lighter and/or $\delta m$ vanishes.
Note that the case with $\ell_{\textrm{lab}}^{\max}=0$ is relevant to not only signals with overlaid vertices, i.e., prompt {\it i}BDM ones, but elastic scattering signals because the latter involves a single interaction point with the absolute lower bound extended down to $m_0 = 30$ MeV.

Since we take a dark photon scenario as in~\eqref{eq:lagrangian}, it is interesting to interpret the parameter reach sensitivities in the usual $m_X$ vs. $\epsilon$ plane, with $m_X$ being the dark photon mass.
We find the minimum value of $\epsilon$ for a given $m_X$ by scanning along the boundary curve in $(\sigma_\epsilon \mathcal{F}, \ell_{\textrm{lab}}^{\max})$, with the mass parameters $m_{0,1,2}$ and coupling constants fixed.
The expected reaches at ProtoDUNE are shown in FIG.~\ref{fig:darkphotonsensitivity} under the assumption that $X$ decays either invisibly (top panels) 
or visibly (bottom panels).
For the former case, we assume $m_X > 2m_1$ in order that the invisible decay modes dominate the visible ones.
Our selection of mass spectra appears in each legend, and $g_{11/12}=1$ for the {\it e}BDM/{\it i}BDM signals in both cases.
The current exclusion limits (brown regions) are extracted from Refs.~\cite{Banerjee:2017hhz} (top)
and \cite{Banerjee:2018vgk} (bottom).
The left panels show the coverage in a worst case scenario assuming 100 background interactions for a one-year exposure and the right panels include the sensitivities in a best case scenario (zero-background assumption) and a two-year exposure of the detector for comparison.

We report the experimental sensitivities for some inelastic scattering scenarios ($\delta m \equiv m_2-m_1 \neq 0$), fixing mass parameters as shown in each legend.
We clearly see that our searches in the {\it i}BDM channels probe parameter regions that are uncovered by existing experimental constraints, by about an order of magnitude in the $\epsilon$ axis depending on the parameter choice.
In the left panels, we also show the results corresponding to different threshold energies, optimistic case 20 MeV (dotted) and pessimistic case 45 MeV (dashed), on top of the baseline value 30 MeV (solid), and find that the coverage in parameter space is mildly affected by the value of energy threshold.
Note that no {\it e}BDM results appear here. The reason is that our estimate for the muon-induced cosmic background to the single-track event is order $10^7 - 10^8$ per year so that sensitivity curves merely lie in the brown regions.
In other words, a suitable cosmic background control should be preceded in order to achieve experimental reach towards unexplored parameter space via {\it e}BDM channels~\cite{Kim:2018veo}.

\section{Conclusions and outlook}

ProtoDUNE possesses an excellent sensitivity to a wide range of BDM parameter space, hence allows a deeper understanding in non-minimal dark-sector physics.
We encourage many theorists to explore phenomenology of their own new physics models at ProtoDUNE.
Moreover, our physics study can be extended to proton scattering and is widely applicable to other existing/future surface-based detectors.

%%%%%%%%%%%%%%%%%%%%%%%%%%%%%%%%%%%%%%%%%%%%%%%%%%%%%%%%%%%%%%%%%%%%%%
\section*{Acknowledgments}
We thank Anne Schukraft for useful discussions, and Gian Giudice for a careful reading of the draft.
DK, JCP and SS also appreciate the hospitality of Fermi National Accelerator Laboratory.
AC and JY are supported by the U.S. Department of Energy, HEP Award DE-SC0011686.  JY thanks the University of Texas at Arlington, the French `Investissements d'avenir' Labex ENIGMASS program and ETH, Zurich for making his stay at CERN possible for this work.
DK is supported by the Korean Research Foundation (KRF) through the CERN-Korea Fellowship program.
JCP is supported by the National Research Foundation of Korea (NRF-2016R1C1B2015225 and NRF-2018R1A4A1025334).
SS is supported by the National Research Foundation of Korea (NRF-2017R1D1A1B03032076).

%%%%%%%%%%%%%%%%%%%%%%%%%%%%%%%%%%%%%%%%%%%%%%%%%%%%%%%%%%%%%%%%%%%%%%

%%%%

\end{document}